\magnification1200

%\rightline{KCL-MTH-14-??}
%\rightline{hep-th/yymmnnn}

\vskip 2cm
\centerline
{\bf  Dual gravity and $E_{11}$ }
\vskip 1cm
\centerline{  Peter West}
\centerline{Department of Mathematics}
\centerline{King's College, London WC2R 2LS, UK}
\vskip 2cm
%\centerline{and}
%\vskip 0.5cm
%\centerline{??,}
%\centerline {??,}
%\centerline{??}
\leftline{\sl Abstract}
 We consider the  equation of motion  in the gravity sector that arises from the non-linear realisation of the semi-direct product of $E_{11}$ and its first fundamental representation  denoted by $l_1$ in four dimensions. This  equation is first order in  derivatives and at  low levels  relates the  usual field of gravity to  a dual gravity field. When the generalised space-time is restricted to be the usual four dimensional space-time we show that this equation does  correctly describe Einstein's theory at the linearised level. We also comment on previous discussions of dual gravity.

\vskip2cm
\noindent

\vskip .5cm

\vfill
\eject

In parts of this paper we will clarify some previous discussions of the dual graviton and as such it will be useful to begin at the beginning.  The dual gravity  field first appeared in  reference [1] which constructed  the equation of motion  of the field $\phi_{ab,c }$ with the symmetries $\phi_{ab,c }= -\phi_{ba, c}$ in a general dimension. The author noted that in the massless case,  and in five dimensions, the equation of motion for this  field had the same number of on-shell degrees of freedom as the graviton. In reference [2] the field $\phi _{a_1\ldots a_{D-3}, b} $ was considered in $D$ dimensions and suggested as a candidate for the dual graviton in the sense that a quantity which contained two space-time derivatives acting on this field could be regarded as being  dual to the Riemann tensor. It was pointed out that when decomposed  to the group SO(D-2) this field contained a component that  had  the same indices as the graviton $h_{a, b}$,  but it was not  shown that the number of degrees of freedom resulting from an analysis of a  field equation were the same as gravity,  or indeed that this field  did provide an alternative description of gravity. 
\par
The next discussion of the dual graviton occurred in reference [3] which was the paper that introduced $E_{11}$. It was shown that at level three the non-linear realisation of $E_{11}$ contained a field of the form $\hat h_{a_1\ldots a_{8}, b} $ and it was suggested that it was the dual of gravity,  generalising the duality between the three and six form fields found at levels one and two respectively. In a separate section  in this paper it was 
it was  shown that  one could rewrite  Einstein's action of gravity  {\bf in a general dimension $D$}  in the form  
$${1\over 2}\int d^D x (\epsilon^{\mu \nu\ldots \tau_{D-2}} 
Y_{\tau_1\ldots \tau_{D-2}}{}^d C_{\mu\nu,d}
+e(-{1\over 2} {(D-3)\over (D-2)}
Y_{\tau_1\ldots \tau_{D-2},}{}^{d} 
Y^{\tau_1\ldots \tau_{D-2},}{}_{d} 
$$
$$+{1\over 2}  (D-2) Y_{\tau_1\ldots\tau_{D-3}d,}{}^{d} 
Y^{\tau_1\ldots\tau_{D-2}e,}{}_{e}  
-{1\over 2}Y_{\tau_1\ldots\tau_{D-3}\kappa,}{}^{d}
Y^{\tau_1\ldots\tau_{D-2} d,}{}_{\kappa}))
\eqno(1)$$
where  
$$ C_{\mu\nu}{}^a= \partial _\mu e_\nu{}^a -\partial _\nu e_\mu{}^a .
\eqno(2)$$
and $Y_{\tau_1\ldots\tau_{D-3}d,}$ is an independent field. 
The  equations of motion for gravity then take the form [3] 
$$ \epsilon^{\mu \tau_1\ldots \tau_{D-1}}\partial_{\tau_1} 
Y_{\tau_2\ldots \tau_{D-1}}{}^d= {\rm {terms\  of\  order}}
(Y_{\tau_1\ldots\tau_{D-2}}{}^d)^2
\eqno(3)$$
and 
$$ \epsilon_{\mu\nu}{}^{\tau_1\ldots \tau_{D-2}}
Y_{\tau_1\ldots \tau_{D-2},b}= -C_{\mu\nu, b}+C_{\nu b, \mu}
-C_{\mu b, \nu}+2(e_{\nu b}C_{\mu c}{}^c -e_{\mu b}C_{\nu c}{}^c )
\eqno(4)$$
\par
Writing $e_\mu{}^a = \delta _\mu^a + h_\mu{}^a$,  and  at lowest order in $h_\mu{}^a$,  the first of these equations implies that 
$$ Y_{\tau_1\ldots \tau_{D-2},b}=\partial _{[ \tau_1}
h_{\tau_2\ldots \tau_{D-2}], b}. 
\eqno(5)$$
Substituting this into equation (4) we find a description of gravity that is  given in terms of the fields $h_\mu^a$ and
$h_{\tau_1\ldots \tau_{D-3},}{}^b$ [3]. Indeed the two fields are related by a first order duality relation.  It is guaranteed by the way that this  field equations was constructed that if we applied another space-time derivative in such a way as to eliminate the field $h_{\tau_1\ldots \tau_{D-3},}{}^b$  then one is left with the field equation for Einstein's linearised theory. At the linearised level, we can use  equation (4) to substitute for $C_{\mu\nu, b}$ in equation (1) to obtain an action entirely in terms of the field $h_{\tau_1\ldots \tau_{D-3},}{}^b$ and 
it was shown in reference [4] that this action, in five dimensions, was the action first proposed in reference [1].
\par
It is important to note that the account of the dual graviton  given in  equations (1-5),   when restricted to eleven dimensions,  was slightly different to that found in the $E_{11}$ non-linear realisation in that   the field $\hat h_{a_1\ldots a_{8}, b} $ found at level three in $E_{11}$ obeys the constraint $\hat h_{[a_1\ldots a_{8}, b] } =0$, while the field $h_{a_1\ldots a_{D-3}, b} $ discussed in equations (1.1-1.5) did not obey this constraint. 
It was observed just  below equation (4.3) in [5]   that equation (4) at the linearised level was only locally Lorentz invariant if the usual transformation of the linearised field of gravity $h_{a,b}$ is accompanied by a transformation of the field $ h_{a_1\ldots a_{D-3}, b} $ of the form 
$$
\delta h_{a,b} =-\Lambda _{ab}  ,\quad \delta  h_{a_1\ldots a_{D-3}, b}= (-1)^D (D-3)! 
\epsilon _{a_1\ldots a_{D-3} b c_1c_2}\Lambda ^{c_1c_2} 
\eqno(6)$$
Thus the local Lorentz transformation affects precisely the part of the field $  h_{a_1\ldots a_{8}, b}$ that is missing in the $E_{11}$ formulation. We will comment further on this point at the end of this paper.   
\par
The fact that the non-linear realisaton of $E_{11}$ contains at level three a field which describes  a dual formulation of gravity provides encouraging evidence for $E_{11}$. However,  it was unclear whether or not the non-linear realisation of $E_{11}$ would lead to precisely  to an  equation of motion for the usual gravity field $h_a{}^b$ and the new dual gravity field $h_{a_1\ldots a_8,b}$ which has the precise form required  to  described gravity. This situation was discussed  in reference [5]. 
Some subsequent papers which discuss the dual graviton are given in [15,16,17]. 
\par
 In order to incorporate space-time in a systematic way it was proposed in 2003 to consider the non-linear realisation of the semi-direct product of $E_{11}$ and its first fundamental representation $l_1$, denoted $E_{11}\otimes_s l_1$ [6]. The reader will be familiar with a semi-direct product as the Poincare group is just the semi-direct product of the Lorentz group and the space-time translations. This construction introduced a generalisation of space-time which  included many more, indeed infinitely many more,  coordinates in addition to the familiar ones. 
The reader who is unfamiliar with  non-linear realisations  and just wants to think about  duality in gravity can skip to the gravity equation of motion 
of equation (15). 
\par
Given any group $G$ with a subgroup $H$,  the non-linear realisation of $G$ with local subgroup $H$ is build from a group element $g\in G$ and it is simply a set of dynamics that is invariant under the transformations 
$$
g\to g_0g \quad {\rm and } \quad g\to gh
\eqno(7)$$
where $g_0\in G$ and $h\in H$ but the $g_0$ transformation is a rigid transformation and the $h$ transformation, like the group element $g\in G$,  is a local transformation. Rigid  means that the transformation does not depend on the space-time in the theory and local means that it does. For some non-linear realisations the space-time is introduced by hand and are in effect dummy variables from the viewpoint of the non-linear realisation but for other non-linear realisations, and this is the case for us, the space-time appears as certain parameters in the group element $g$. The method of non-linear realisations is  precisely defined  and,  once the number of space-time derivative is specified,  the resulting dynamics is unique,  or at worst determined up to a number of constants. 
\par
We wish to study  the non-linear realisation of the group $E_{11}\otimes_s l_1$ with the local subgroup being  the Cartan involution invariant subalgebra of $E_{11}$.   The fields of the theory  appear in the part of the group element $g$ associated with   $E_{11}$,  while the parameters in the part of the group element $g$ associated with $l_1$ are the coordinates of the generalised space-time. For a review of non-linear realisations and the non-linear realisation of $E_{11}\otimes_s l_1$ see reference [7]. 
\par
While fragments  of the non-linear realisation of $E_{11}\otimes_s l_1$ have been previously worked out, see  for example [3]   [8] and  [9], it is only in relatively recently that this non-linear realisation has  systematically been computed first in eleven dimensions [10] and then in four dimensions [11,12] at low levels.  As is appropriate for  theories invariant under duality symmetries it has been assumed in these references  that the equations of motion of the $E_{11}\otimes_s l_1$ non-linear realisation are first order in the generalised space-time derivatives and as a result they can be constructed from the Cartan forms which only transform under the local subgroup.  
In this paper we will not need to understand in detail how the non-linear realisation works, but we will need the resulting dynamics that involves  the gravity sector found in the four dimensional theory.  

The four dimensional theory is found by decomposing $E_{11}\otimes _s l_1$ with respect to the subgroup $GL(4)\otimes E_7$ which results from deleting node four in the usual labelling  of the Dynkin diagram of $E_{11}$. The $GL(4)$ factor is associated with gravity and its indices are those of the usual four dimensional space-time. 
In four dimensions the fields (generators) can be classified according to a level which is just the number of down (up) minus up (down)  SL(4) indices that they carry. The subalgebra $GL(4)\otimes E_7$  is the algebra of $E_{11}$ at level zero and the Cartan  involution invariant subalgebra of $E_{11}$ at  level zero is $SO(4)\otimes SU(8)$. The fields in the four dimensional theory at lowest levels are 
$$
h_a{}^b (0), \phi^\alpha (0) , A_{aN} (1), \hat h_{ab} (2), \ldots 
\eqno(8)$$
where the number in the bracket is the level, the indices $a,b=1,\ldots , 4$ and the label $N$ takes the values $N=1,\dots , 56$. The first entry is the graviton, the vielbein being given by $e_\mu{}^a = (e^h) _\mu{}^a$,  the second entry contains  the scalars which belong to the non-linear realisation of $E_7$ with local subgroup SU(8), the third entry contains the gauge fields which belong to the 56-dimensional representation of $E_7$ and the final  entry contains the dual graviton which obeys $\hat h_{ab}=\hat  h_{ba}$, that is,  it satisfies the previously mentioned constraint $\hat h_{[a,b]}=0$. 
\par
The generalised space-time contains the usual coordinate $x^a$ and at the next level the coordinates $z^N$ which belongs to the 56-dimensional representation of $E_7$ [6,13,14] 
as well as unfamiliar coordinates at all higher levels. Reference [11]  considered   the  equations of motion that result from the non-linear realisation of $E_{11}\otimes _s l_1$ when we keep only the fields of equation (8) as well as  the usual coordinates of space-time $x^a$ and the coordinates $z^N$. With this restriction the  non-linear realisation of $E_{11}\otimes _s l_1$ leads to unique  equations of motion for   the vectors and the scalars [11,12]  which agree precisely with those of maximal supergravity in four dimensions when we further restrict to a space-time that is the usual one with coordinates $x^\mu$. When carrying out this calculation the local subalgebra was used to gauge away as much of the group element $g$ as possible, except at level zero, that is, the local symmetry $SO(4)\otimes SU(8)$ was left unfixed. To construct the equations of motion without this symmetry being manifest would be very complicated as even Lorentz symmetry would not be manifest.  
\par
Beginning in the   appendix of reference [11] and  computed in more detail  in a to be published paper [12], the variation of the vector equation under the $E_{11}$ symmetries, was found to lead to an equation relating the graviton to the dual graviton which, when  restricted to contain just these fields and to the usual coordinates of space-time, is given by 
$$
E_{a_1a_2 ,} {}_c\equiv P_{a_1,a_2 a_3} - P_{a_2, a_1 a_3} 
-2\chi (Q_{a_1, a_2 a_3} - Q_{a_2, a_1 a_3})
$$
$$- \epsilon _{a_1a_2b_1b_2} \hat P_{b_1, b_2 a_3} 
-{1\over 2} \epsilon _{a_1a_2a_3b} \hat  P_{b, c} {}^c=0 
\eqno(9)$$
where 
$$
P_{a,bc}= e_a{}^\mu e_{(b|}{}^\tau \partial_\mu e_\tau{}_{|c)}, \quad 
Q_{a,bc}= e_a{}^\mu e_{[b|}{}^\tau \partial_\mu e_\tau{}_{|c]}, \quad
\eqno(10)$$
and 
$$
\hat P_{a, bc}= e_a{}^\mu ( \partial_\mu \hat h_{b_1b_2} + (e^{-1} \partial_\mu e)_{b_1}{}^{c} \hat h_{cb_2}+ (e^{-1} \partial_\mu e)_{b_2}{}^{c} \hat h_{b_1 c})= e_a{}^\mu e_{b_1}{}^{\rho_1} e_{b_2}{}^{\rho_2} 
\partial_\mu \hat h_{\rho_1\rho_2}
\eqno(11)$$
The objects denoted by the symbols $P_{a,bc} $ and $Q _{a,bc} $ are   Cartan forms of $E_{11}$ which can be divided into those that  
are odd ($P$) and even ($Q)$ respectively under the relevant part of the Cartan involution which is just SO(4) for these objects.  The value of the non-zero constant $\chi$,  and indeed the equation itself, is the subject of further study [12] which depends on the analysis of the terms that contain derivatives with respect to   the higher level coordinates. Nonetheless  this equation is  sufficiently  well known for the purposes of this paper. 

\par
The non-linear realisation of $E_{11}\otimes_s l_1$ has some unusual features  compared to the more familiar cases of non-linear realisations. In particular the equations are not just constructed from the odd Cartan forms, in this case,  
$P_{a,bc}$ and $\hat P_{a,bc}$ but also  the even Cartan forms $Q_{a,bc}$. This is at first sight unexpected as in the general theory of non-linear realisations the Cartan forms are inert under the rigid transformation of equation (7) but under the local transformations 
the odd and even parts of the Cartan forms transform into each other separately. However, once one fixes part of the group element using the local symmetry, as one usual does, then the corresponding parts of the even and odd Cartan forms associated with the local transformations used to fix the group become equal. In our case this means that the even and odd Cartan forms are all equal except  for those at level zero, that is the ones in equation (10) and 
their $E_7$ analogues which do not concern us here. As a result the variation of the even Cartan forms, that is, the $Q$'s leads to other $Q$'s but these are equal, if they are above level zero, to the $P$'s and so can cancel terms that arise from the variation of other $P$'s. Indeed, it turns out that only by this mechanism can some level one terms be cancelled. 
\par
This subtle, but crucial,  point has  arisen for essentially the first time in the context of the $E_{11}\otimes _s l_1$ non-linear realisation and it is associated with the different  treatment of  the fields at level zero  to those  at higher levels. This origin of this difference is that, as we have mentioned, the group element $g$ of equation (7) has been fixed using the local transformation at all levels except level zero. The above gravity equation was found by varying the  
the scalar and  vector  equations of motion under the local $E_{11}$ parameter that involves the generators at level one and minus one; it has a parameter of the form   $\Lambda_{a_1a_2a_3}$. This  variation can only be cancelled  if one adds to the gravity equation terms that contain the level zero Cartan odd forms, that is the $Q_a{}_{,bc}$ that occur in  bracket in the first line of equation (9). Indeed these terms vary into level one Cartan forms that are odd but,  as just explained, since these are equal to  the even level one Cartan forms they  can be used to cancel variations that arise from other  even Cartan forms. This point will be further explained in reference [12].  
\par
The presence of the Cartan forms $Q_{a,bc}$ has an important consequence for the symmetries of the equation of motion. Under a local Lorentz transformation  the objects  $P_{a,bc}$ transform covariantly under the local subgroup while the $Q_{a,bc}$ transform inhomogeneously; in particular under a local Lorentz transformation one finds that 
$$
\delta  P_{a,bc}= \Lambda _a {}^e P_{e,bc}+ \Lambda _b {}^e P_{a,e c}+ \Lambda _c {}^e P_{a,be}
$$
$$ \delta  \hat P_{a,bc}= \Lambda _a {}^e \hat P_{e,bc}+ \Lambda _b {}^e\hat  P_{a,e c}+ \Lambda _c {}^e \hat P_{a,be}
\eqno(12)$$
while 
$$
\delta Q_{a,bc}= \Lambda _a {}^e Q_{e,bc}+ \Lambda _b {}^e Q_{a,e c}+ \Lambda _c {}^e Q_{a,be}- e_a{}^\mu \partial_\mu  \Lambda _{bc} 
\eqno(13)$$
The reader can easily verify that these are the transformation that result form making a local Lorentz transformation of the vierbein $e_\mu{}^a\to \Lambda^a{}_b e_\mu {}^b$ in the usual way. 
\par
It follows that as the gravity equation (9)  contains the Cartan forms $Q_{a,bc}$ it   can not be Lorentz invariant. Indeed, under a local Lorentz transformation $E_{a_1a_2} {}_{,c}$ transforms as 
$$
\delta E_{a_1a_2} {}_{,c}= \Lambda_{a_1}{}^ d E_{ d a_2} {}_{,c}
+\Lambda_{a_2}{}^ d E_{ a_1d } {}_{,c}+ \Lambda_{c}{}^ d E_{ a_1 a_2} {}_{,d}+2\chi(e_{a_1}{}^\mu \partial_\mu  \Lambda _{a_2a_3}
-e_{a_2}{}^\mu \partial_\mu  \Lambda _{a_1a_3})
\eqno(14)$$
Hence, although the gravity equation  was derived by demanding invariance under a  transformation involving level one and minus one generators the resulting equation is not invariant under the level zero  local Lorentz transformations. The correct interpretation is to regard the gravity equation (9) to hold  only 
 modulo the above local Lorentz transformations. Put another way 
 as $E_{a_1a_2} {}_{,c}$  vanishes up to Lorentz transformations which can be stated in the form 
$$
E_{a_1a_2} {}_{,c} -4\chi(e_{a_1}{}^\mu \partial_\mu  \Lambda _{a_2a_3}
-e_{a_2}{}^\mu \partial_\mu  \Lambda _{a_1a_3})=
P_{a_1,a_2 a_3} - P_{a_2, a_1 a_3} 
-2\chi (Q_{a_1, a_2 a_3} - Q_{a_2, a_1 a_3})
$$
$$- \epsilon _{a_1a_2b_1b_2} \hat P_{b_1, b_2 a_3} 
-{1\over 2} \epsilon _{a_1a_2a_3b} \hat  P_{b, c} {}^c
+2\chi(e_{a_1}{}^\mu \partial_\mu  \Lambda _{a_2a_3}
-e_{a_2}{}^\mu \partial_\mu  \Lambda _{a_1a_3})=0
\eqno(15)$$
\par
The usual spin connection $\omega_{\mu ,} {}_{a}{}^{b}$ is given in terms of these quantities by 
$$
\omega_{c,} {}_{a b}= e_c{}^\mu\omega_{\mu ,}{}_{a b}= - P_{a,bc} + P_{b,ac} + Q_{c,ab} 
\eqno(16)$$
Thus we can rewrite equation (15) in the form 
$$
- \omega _{a_3,a_1a_2} + Q_{a_3,a_1a_2} -4\chi( Q_{a_1,a_2 a_3} - Q_{a_2, a_1 a_3}) -\epsilon _{a_1a_2b_1b_2} \hat P_{b_1, b_2a_3}
-{1\over 2} \epsilon_{a_1a_2a_3 d}\hat P_{d, c}{}^c
$$
$$
+2\chi(e_{a_1}{}^\mu \partial_\mu  \Lambda _{a_2a_3}
-e_{a_2}{}^\mu \partial_\mu  \Lambda _{a_1a_3})=0
\eqno(17)$$
\par
In this paper we will analyse the  equation that the non-linear realisation of $E_{11}\otimes l_1$ predicts for gravity, that is equation (15),  or equivalently equation (17), at the linearised level. Taking $e_\mu{}^a = \delta _\mu ^a + h_\mu {}^a $ we will work to linear order in $h_\mu {}^a $. In this approximation 
$$
P_{a,bc}= \partial_a h_{(bc)}, \quad 
Q_{a,bc}=\partial_a h_{[ bc]}, \quad \hat P_{a,bc}= \partial_a \hat h_{bc}
\eqno(18)$$
 whereupon equation (15) becomes 
$$
E_{a_1a_2, a_3}^L\equiv 
\partial_{a_1} h_{(a_2a_3)}- \partial_{a_2} h_{(a_1a_3)}
- \epsilon_{a_1a_2b_1b_2} \partial_{b_1}\hat h_{b_2a_3} 
-{1\over 2} \epsilon_{a_1a_2a_3b} \partial^{b}\hat h_{c}{}^c 
$$
$$
-2\chi (\partial_{a_1} h_{[a_2a_3]}- \partial_{a_2} h_{[a_1a_3]}
-\partial_{a_1} \Lambda_{[a_2a_3]}+ \partial_{a_2} \Lambda_{[a_1a_3]})=0 
\eqno(19)$$
\par
To find the equation for gravity in terms of the usual field $h_{ab}$ we must manipulate  equation (19) so that it no longer contains the dual gravity 
$\hat h_{ab}$ and also eliminate the local Lorentz transformation. We can eliminate the dual gravity field in two ways; we can act with $\partial^{a_2}$ to find 
$$
Z_{a_1a_3}\equiv \partial^{a_2} E_{a_2a_2, a_3}= 
\partial_{a_1}\partial^{a_2} h_{(a_2a_3)}- \partial^2h_{(a_1a_3 )}  
-\partial_{a_1} \partial^{a_2} k_{a_2a_3} + \partial^2 k_{a_1a_3}=0 
\eqno(20)$$
where $\partial^2= \partial^c\partial_c$, and $k_{ab}=2\chi  (h_{[ab]}- \Lambda _{ab} )$, or  we can also trace on $a_2$ and $a_3$ to find 
$$
\partial_{a_1} h^b{}_b - \partial^b h_{(a_1 b)} + \partial_b k_{a_1}{}^b=0
\eqno(21)$$
To eliminate the local Lorentz transformation we can act on equation (21) with $\partial_{a_3}$,  subtract the result from equation (20),  and then symmetrise  with respect to $a_1$ and $a_3$. The final equation which is independent of the dual graviton and the local Lorentz transformation  is 
$$
 \partial _{a_1} \partial^b h_{(ba_3)}+ \partial_{a_3} \partial^b h_{(ba_1)} - \partial^2 h_{(a_1a_3)} - \partial_{a_3 }\partial_{a_1} h_b{}^b=0
\eqno(22)$$
This is indeed the linearised Einstein equation.
\par
Some discussions of dual gravity have focused on the fact that the dual graviton field which comes from the $E_{11}$ non-linear realisation is subject to the constraint mentioned above, which in $D$ dimension is given by $\hat h_{[a_1\ldots a_{D-3} ,b]}=0$. In particular, they considered that  this constraint may lead to problems in finding the correct equation of motion for gravity. However, as it apparent from above this constraint has not prevented us from deriving the correct linearised Einstein equation.  In view of the confusion that surrounds this issue we will now discuss it in more detail. The apparent problem can be stated as follows; the non-linear realisation results in an equation that involves three indices and contains the  graviton and the dual graviton. However, due to the constraint once one traces on two indices one finds an equation that has one index and no dual graviton. Indeed it is straightforward to verify that  if the dual graviton obeys the constraint 
$\hat h_{[a_1\ldots a_{D-3},b]}=0$ then there is no possible term involving epsilon one space-time derivative and this field. The resulting equation has   one Lorentz index, one space-time derivative and  involves the graviton field alone.  However, such an equation is not  compatible with  Einstein's equations at the linearised level. For example, in  four  dimensions upon tracing equation (19) we find equation (21), which in  the absence of  the realisation that the equation of motion only holds up to Lorentz transformations, results in the equation  
$$
\partial_{a} h^b{}_b - \partial^b h_{(a b)} =0
\eqno(23)$$
which is clearly not an equation belonging to  Einstein's gravity. 
 However, due to  the presence of the 
 Cartan form $Q_{a,bc}$ we have an   associated local Lorentz transformation  and then we   find equation (21) which includes a local Lorentz transformation. As we have explained above to  find an equation that is independent of the local Lorentz transformation and so having taken the trace we must then act with $\partial ^{a_1}$ to find the equation 
$$
\partial^2 h^b{}_b - \partial ^{a} \partial^b h_{a b}=0
\eqno(24)$$
which is indeed the linearised version of the trace of the Einstein equation. 
\par
As we noted earlier,  the discussion of reference   [3] in $D$ dimensions  involved the  dual graviton field $h_{a_1\ldots a_{D-3}, b} $ which was subject to  the above constraint and by its derivation did describe linearlised gravity correctly.   However, as we discussed,  the totally antisymmetric part of the dual graviton $h_{[a_1\ldots a_{D-3}, b]} $ could be removed in equation (4) by carrying out a Lorentz transformation as in equation (6) [5]. Hence one   arrives at a formulation that has a  dual gravity field that is  constrained at the cost of having an equation that was modulo a local Lorentz equation.  This formulation differs from the one considered in the bulk of this paper in that it has different trace of the spin connection terms. Since the appearance of the Cartan form $Q_{a,bc}$ in the gravity equation determines which  local Lorentz transformations  the equation is  modulo there is some freedom to formulate  the equation and still have it describe gravity at the linearised level. We note that taking the equation of gravity to be first order in space-time derivatives, but to hold modulo local Lorentz transformations,  is a new way of formulating the equation of gravity. 
\par
We now comment on the discussion in reference [10] which, in a section added,  gave the equation of motion of the $E_{11}\otimes _s l_1$ non-linear realisation in the gravity sector in eleven dimensions. It was realised in this paper that this equation should   hold only modulo Lorentz transformations,  however, what the author of this reference did not appreciate was that if one did take this into account then one found the correct linearised Einstein equation and as a result a number of the comments in the note added to  this paper are misleading. 
\par
In this paper we have considered  the equation of motion that follows from the  $E_{11}\otimes_s l_1$ non-linear realisation in the gravity sector. Although this equation is still subject to further study we think its form as given in equation (9) is correct. This equation  expresses the duality between the usual graviton field and the dual graviton and we have shown that it leads to the correct equation of motion, that is, Einstein's equation, at the linearised level. Although this result was shown in four dimensions one can be confident it will hold in all dimensions for which theories arise in the 
$E_{11}\otimes_s l_1$ non-linear realisation. We recall that this non-linear realisation leads to the correct equations for the scalars and the form fields and so to show the $E_{11}$ conjecture one just has to show that it leads not only to  Einstein equation at the  linearised level, but to the full non-linear equation. Work towards this goal continues [12].  
\par
Although the non-linear realisation process is simple to state when carrying it out one encounters a number of rather subtile points, some of which we have discussed above. We think that the pattern encountered for the gravity equation in the $E_{11}\otimes_s l_1$ non-linear realisation is the generic one, meaning that the  equations which exhibit the symmetries are first order in derivatives of the generalised space-time, but they only hold modulo certain local symmetries. To eliminate the latter one needs to take higher derivatives and so find the dynamical equations. The number of derivatives will increase as the level increases corresponding to the fact that the fields at higher levels have mixed index structures of increasing complexity. It follows from this that a Riemannian approach to the geometry, although it may work at very low levels, is unlikely to be very useful in the full theory.

%%%%%%%%%%%%%%%%%%%%%%%%%%%%%%%%%%%%%%%%%%%%%%%%%%%%%%%%%%%%%%%%%
\medskip
{\bf {Acknowledgment}}
\medskip 
I wish to thank Nikolay Gromov and Dmitri Sorokin for discussions and the SFTC for support from Consolidated grant number ST/J002798/1.

%%%%%%%%%%%%%%%%%%%%%%%%%%%%%%%%%%%%%%%%%%%%%%%%%%%%%%%%%%%%%

\medskip
{\bf {References}}
\medskip
\item{[1]} T. Curtright, {\it Generalised Gauge fields}, Phys. Lett. {\bf 165B} (1985) 304. 
\item{[2]} C. Hull, {\it   Strongly Coupled Gravity and Duality}, Nucl.Phys. {\bf B583} (2000) 237, hep-th/0004195. 
\item{[3]} P. West, {\it $E_{11}$ and M Theory}, Class.Quant.Grav. 18 (2001) 4443-4460, hep-th/0104081. 
\item{[4]} N.  Boulanger, S. Cnockaert  and  M.  Henneaux, {\it A note on spin-s duality}, JHEP 0306 (2003) 060, hep-th/0306023. 
\item{[5]}  P. West, {\it Very Extended $E_8$ and $A_8$ at low levels, Gravity and Supergravity}, Class.\ Quant.Grav. 20 (2003) 2393-2406, hep-th/0212291.
   \item{[6]} P. West, {\it $E_{11}$, SL(32) and Central Charges},
Phys. Lett. {\bf B 575} (2003) 333-342, {\tt hep-th/0307098}
\item {[7]} P. West, {\it Introduction to Strings and Branes}, Cambridge University Press 2012. 
\item{[8]} I. Schnakenburg and P. West, {\it Kac-Moody Symmetries of IIB Supergravity}, Phys.Lett. B517 (2001) 421-428, hep-th/0107181. 
\item{[9]}  ÊF. Riccioni and P. West, {\it E(11)-extended spacetime and gauged supergravities},
JHEP {\bf 0802} (2008) 039, ÊarXiv:0712.1795
\item{[10]}  P. West, {\it Generalised geometry, eleven dimensions and E11}, 
JHEP 1202 (2012) 018, arXiv:1111.1642. 
\item{[11]}  P. West, {\it E11, Generalised space-time and equations of motion in four dimensions}, JHEP 1212 (2012) 068, arXiv:1206.7045. 
\item{[12]}  N. Gromov A. Tumanov and  P. West to appear. 
\item{[13]} P. West,  {\it $E_{11}$ origin of Brane charges and U-duality
multiplets}, JHEP 0408 (2004) 052, hep-th/0406150. 
\item{[14]} P. West, {\it Brane dynamics, central charges and
$E_{11}$}, hep-th/0412336. 
\item{[15]} N. Boulanger and O. Holm, {\it Non-linear parent action and dual gravity }, Phys.Rev.D78 (2008) 064027,  arXiv:0806.2775. 
\item{[16]} E. Bergshoeff, M de Roo, S. Kerstan, A.
Kleinschmidt, F. Riccioni, {\it  Dual Gravity and Matter}, 
   Gen.Rel.Grav.41 (2009) 39-48, arXiv:0803.1963.  
\item{[17]} E. Bergshoeff, M de Roo, O, Hohm, {\it  Can dual gravity be
reconciled with E11?},  Phys.Lett.{\bf B675} (2009) 371-376,
arXiv:0903.4384.  

%%%%%%%%%%%%%%%%%%%%%%%%%%%%%%%%%%%%%%%%%%%%%%%%%

\end